\documentclass[10pt,twoside,twocolumn]{IEEEtran}
\newcommand\nc\newcommand
\usepackage{times,amssymb,amsmath,epsfig,nicefrac,euscript,cite,mathrsfs,balance}
\usepackage{float}
\newtheorem{theorem}{Theorem}[section]

\newtheorem{lemma}[theorem]{Lemma}
\newtheorem{proposition}[theorem]{Proposition}

\newtheorem{definition}[theorem]{Definition}

\newcommand{\ol}{\setlength{\itemsep}{-40pt}\begin{enumerate}}
\newcommand{\eol}{\end{enumerate}\setlength{\itemsep}{-\parsep}}

\nc\bfa{{\boldsymbol a}}\nc\bfA{{\bf A}}\nc\cA{{\mathcal A}}\nc\sA{{\mathscr A}}
\nc\bfb{{\boldsymbol b}}\nc\bfB{{\bf B}}\nc\cB{{\mathcal B}}\nc\sB{{\mathscr B}}
\nc\bfc{{\boldsymbol c}}\nc\bfC{{\bf C}}\nc\cC{{\mathcal C}}\nc\sC{{\mathscr C}}
\nc\bfd{{\boldsymbol d}}\nc\bfD{{\bf D}}\nc\cD{{\mathcal D}}\nc\sD{{\mathscr D}}
\nc\bfe{{\boldsymbol e}}\nc\bfE{{\bf E}}\nc\cE{{\mathcal E}}\nc\sE{{\mathscr E}}
\nc\bff{{\boldsymbol f}}\nc\bfF{{\bf F}}\nc\cF{{\mathcal F}}\nc\sF{{\mathscr F}}
\nc\bfg{{\boldsymbol g}}\nc\bfG{{\bf G}}\nc\cG{{\mathcal G}}\nc\sG{{\mathscr G}}
\nc\bfh{{\boldsymbol h}}\nc\bfH{{\bf H}}\nc\cH{{\mathcal H}}\nc\sH{{\mathscr H}}
\nc\bfi{{\boldsymbol i}}\nc\bfI{{\bf I}}\nc\cI{{\mathcal I}}\nc\sI{{\mathscr I}}
\nc\bfj{{\boldsymbol j}}\nc\bfJ{{\bf J}}\nc\cJ{{\mathcal J}}\nc\sJ{{\mathscr J}}
\nc\bfk{{\boldsymbol k}}\nc\bfK{{\bf K}}\nc\cK{{\mathcal K}}\nc\sK{{\mathscr K}}
\nc\bfl{{\boldsymbol l}}\nc\bfL{{\bf L}}\nc\cL{{\mathcal L}}\nc\sL{{\mathscr L}}
\nc\bfm{{\boldsymbol m}}\nc\bfM{{\bf M}}\nc\cM{{\mathcal M}}\nc\sM{{\mathscr M}}
\nc\bfn{{\boldsymbol n}}\nc\bfN{{\bf N}}\nc\cN{{\mathcal N}}\nc\sN{{\mathscr N}}
\nc\bfo{{\boldsymbol o}}\nc\bfO{{\bf O}}\nc\cO{{\mathcal O}}\nc\sO{{\mathscr O}}
\nc\bfp{{\boldsymbol p}}\nc\bfP{{\bf P}}\nc\cP{{\mathcal P}}\nc\sP{{\mathscr P}}
\nc\bfq{{\boldsymbol q}}\nc\bfQ{{\bf Q}}\nc\cQ{{\mathcal Q}}\nc\sQ{{\mathscr Q}}
\nc\bfr{{\boldsymbol r}}\nc\bfR{{\bf R}}\nc\cR{{\mathcal R}}\nc\sR{{\mathscr R}}
\nc\bfs{{\boldsymbol s}}\nc\bfS{{\bf S}}\nc\cS{{\mathcal S}}\nc\sS{{\mathscr S}}
\nc\bft{{\boldsymbol t}}\nc\bfT{{\bf T}}\nc\cT{{\mathcal T}}\nc\sT{{\mathscr T}}
\nc\bfu{{\boldsymbol u}}\nc\bfU{{\bf U}}\nc\cU{{\mathcal U}}\nc\sU{{\mathscr U}}
\nc\bfv{{\boldsymbol v}}\nc\bfV{{\bf V}}\nc\cV{{\mathcal V}}\nc\sV{{\mathscr V}}
\nc\bfw{{\boldsymbol w}}\nc\bfW{{\bf W}}\nc\cW{{\mathcal W}}\nc\sW{{\mathscr W}}
\nc\bfx{{\boldsymbol x}}\nc\bfX{{\bf Z}}\nc\cX{{\mathcal X}}\nc\sX{{\mathscr X}}
\nc\bfy{{\boldsymbol y}}\nc\bfY{{\bf Y}}\nc\cY{{\mathcal Y}}\nc\sY{{\mathscr Y}}
\nc\bfz{{\boldsymbol z}}\nc\bfZ{{\bf Z}}\nc\cZ{{\mathcal Z}}\nc\sZ{{\mathscr Z}}
\nc\bsa{{\boldsymbol a}}\nc\bsA{{\boldsymbol A}}
\nc\bsb{{\boldsymbol b}}\nc\bsB{{\boldsymbol B}}
\nc\bsc{{\boldsymbol c}}\nc\bsC{{\boldsymbol C}}
\nc\bsd{{\boldsymbol d}}\nc\bsD{{\boldsymbol D}}
\nc\bse{{\boldsymbol e}}\nc\bsE{{\boldsymbol E}}
\nc\bsf{{\boldsymbol f}}\nc\bsF{{\boldsymbol F}}
\nc\bsg{{\boldsymbol g}}\nc\bsG{{\boldsymbol G}}
\nc\bsh{{\boldsymbol h}}\nc\bsH{{\boldsymbol H}}
\nc\bsi{{\boldsymbol i}}\nc\bsI{{\boldsymbol I}}
\nc\bsj{{\boldsymbol j}}\nc\bsJ{{\boldsymbol J}}
\nc\bsk{{\boldsymbol k}}\nc\bsK{{\boldsymbol K}}
\nc\bsl{{\boldsymbol l}}\nc\bsL{{\boldsymbol L}}
\nc\bsm{{\boldsymbol m}}\nc\bsM{{\boldsymbol M}}
\nc\bsn{{\boldsymbol n}}\nc\bsN{{\boldsymbol N}}
\nc\bso{{\boldsymbol o}}\nc\bsO{{\boldsymbol O}}
\nc\bsp{{\boldsymbol p}}\nc\bsP{{\boldsymbol P}}
\nc\bsq{{\boldsymbol q}}\nc\bsQ{{\boldsymbol Q}}
\nc\bsr{{\boldsymbol r}}\nc\bsR{{\boldsymbol R}}
\nc\bss{{\boldsymbol s}}\nc\bsS{{\boldsymbol S}}
\nc\bst{{\boldsymbol t}}\nc\bsT{{\boldsymbol T}}
\nc\bsu{{\boldsymbol u}}\nc\bsU{{\boldsymbol U}}
\nc\bsv{{\boldsymbol v}}\nc\bsV{{\boldsymbol V}}
\nc\bsw{{\boldsymbol w}}\nc\bsW{{\boldsymbol W}}
\nc\bsx{{\boldsymbol x}}\nc\bsX{{\boldsymbol X}}
\nc\bsy{{\boldsymbol y}}\nc\bsY{{\boldsymbol Y}}
\nc\bsz{{\boldsymbol z}}\nc\bsZ{{\boldsymbol Z}}

\newcommand{\bfit}{\bfseries\itshape}

\newcommand\integers{{\mathbb Z}}

\nc\uz{\underline z}

\newcommand{\remove}[1]{}
\nc{\FS}{{\mathfrak S_n}}
\nc\half{\nicefrac12}
\begin{document}

% paper title
\title{Codes in Permutations and Error Correction for Rank Modulation} \author{Alexander~Barg~and~Arya
  Mazumdar \thanks{Alexander Barg is with the Department of Electrical
    and Computer Engineering and Institute for Systems Research,
    University of Maryland, College Park, MD 20742 and Institute for
    Problems of Information Transmission, Moscow, Russia (e-mail:
    abarg@umd.edu).}  \thanks{Arya Mazumdar is with the Department of
    Electrical and Computer Engineering and Institute for Systems
    Research, University of Maryland, College Park, MD 20742 (e-mail:
    arya@umd.edu).}  \thanks{Research supported in part by NSF grants
    CCF0830699, CCF0635271, DMS0807411.} }

\maketitle

\begin{abstract}
  Codes for rank modulation have been recently proposed as a means of
  protecting flash memory devices from errors. We study basic coding
  theoretic problems for such codes, representing them as subsets of
  the set of permutations of $n$ elements equipped with the Kendall
  tau distance. We derive several lower and upper bounds on the size of
  codes. These bounds enable us to establish the exact scaling of
  the size of optimal codes for large values of $n$. We also
  show the existence of codes whose size is within a constant factor of
  the sphere packing bound for any fixed number of errors.
\end{abstract}

{\bfit Index terms}---{\small\bf Bose-Chowla theorem, flash memory, inversion,
Kendall tau distance, rank permutation codes.}

\section{Introduction}
Codes in permutations form a classical subject of coding
theory. Various metric functions on the symmetric group $\FS$ have
been considered, giving rise to diverse combinatorial problems. The
most frequently studied metric on $\FS$ is the Hamming distance. Codes
in $\FS$ with the Hamming distance, traditionally called permutation
arrays, have been a subject of a large number of papers; see, e.g.,
the works by Blake et al. \cite{bla79} and Colbourn et
al. \cite{col04}.

In this paper we are interested in a different metric on $\FS$ which
we proceed to define.  Let $\sigma=(\sigma(1),\dots,\sigma(n))$ be a
permutation of the set $[n]=\{1,2,\dots, n\}.$ The {\em Kendall
tau  distance} $d_\tau(\sigma,\pi)$ from $\sigma$ to another permutation
$\pi$ is defined as the minimum number of transpositions of pairwise
adjacent elements required to change $\sigma$ into $\pi.$ Denote by
$X_n=(\FS,d_\tau)$ the metric space of permutations on $n$ elements
equipped with the distance $d_\tau.$

The Kendall distance originates in statistics and has been adopted as
a measure of quality of codes under the so-called rank modulation
scheme first considered by Chadwick and Kurz \cite{CK1969}. In this
scheme, the transmitted sequences are given by permutations of $n$
elements while information is carried by the relative magnitude (rank)
of elements in the permutation rather than by the absolute value of
the elements. The motivation for considering this scheme in
\cite{CK1969} stems from systems in which transmitted signals are
subjected to impulse noise that changes the value of the signal
substantially but has less effect on the relative magnitude of the
neighboring signals.  Recently (and independently of \cite{CK1969})
rank modulation was suggested by Jiang et al. \cite{JSB2008,jia08b} as
a means of efficient writing of information into flash memories.
Rewriting the contents of a group of memory cells is easy if one needs
to increase the charges of the cells or leave some of them unchanged
and impractical if some of the charges need to be decreased.
Furthermore, reliability of the data stored in
flash memory is affected by the drift in the charge of the cells
caused for instance by aging devices or other reasons. Since the drift
in different cells may occur at different speed, errors introduced in
the data are adequately accounted for by tracking the relative value of
adjacent cells, i.e., the Kendall distance between the groups of cells in memory.
These considerations make
rank modulation suitable for coding for flash memories. More details
of both the writing and the error processes in memory are given in
\cite{JSB2008} and references in that paper.

The focus of our work is on bounds and constructions of codes in the
Kendall space $X_n.$ Coding-theoretic considerations call for
estimating the volume of the sphere in $X_n$ because it can be used to
derive basic bounds on the size of codes. Spheres in the Kendall
space were studied by analytic means in a number of earlier works
\cite{M2001,lou03} relying on the well-known correspondence of
permutations and their inversion vectors; however it turned out that
code bounds that can be obtained from these works do not lead to
nontrivial (other than 0 or 1) values of the code rate.  Regarding
specific code families for correcting Kendall errors, the only
previous work is that by Jiang et al. \cite{JSB2008} who constructed a
family of single-error-correcting codes of size $M\ge \frac12(n-1)!,$
i.e., at least half the maximum possible.

\vspace*{1mm} {\em Our results.}  In this paper we discuss several
possible ways to bound the size of codes for rank modulation of a
given distance, often calling them rank permutation codes.
We derive a Singleton-type bound and sphere-packing
bounds on such codes. Since the maximum value
of the distance in $X_n$ is $\binom n2$, this leaves a number
of possibilities for the scaling rate of the distance for asymptotic
analysis, ranging from $d=O(n)$ to $d=\Theta(n^2).$
These turn out to be the two extremes for the size of optimal
rank permutation codes. Namely, earlier work in combinatorics of permutations
implies that a code with distance $d=\Theta(n^2)$ occupies a vanishing
proportion of the space $X_n$ while a code of distance $O(n)$
can take a close-to-one proportion of its volume.
We cover the intermediate cases,
showing that the size of optimal codes with distance
$d\sim n^{1+\epsilon}, 0<\epsilon<1$ scales as $\exp((1-\epsilon)n\ln n).$
It is interesting that unlike many other asymptotic coding problems,
the Kendall space of permutations affords an exact answer for
the growth rate of the size of optimal codes.
The proof of the bounds relies on weight-preserving embeddings
of $X_n$ into other metric spaces which provide insights into the
asymptotic size of codes.

We also show the existence of a family of rank permutation codes that
correct a
constant number of errors and have size within a constant factor of
the sphere packing bound. The construction relies on the well-known
Bose-Chowla Theorem in additive number theory.

Section II of our paper is devoted to the relation of the Kendall
metric space to other metric spaces related to permutations.
In Section III we use these insights to derive bounds on
codes for rank modulation,
and conduct their asymptotic analysis. Section IV contains
a construction of $t$-error-correcting rank permutation codes.

\section{Weight-preserving embeddings of the Kendall metric space}
We begin with recalling basic properties of the distance $d_\tau$ such
as its relation to the number of inversions in the permutation, and
weight-preserving embeddings of $\FS$ into other metric spaces. Their
proofs and a detailed discussion are found for instance in the books
by Comtet \cite{com74} or Knuth \cite[Sect. 5.1.1]{K1973}.
%\cite[pp.43ff]{bon04}, Comtet \cite[pp.236-240]{com74}, or Knuth \cite[p.15]{K1973}.

The distance $d_\tau$ is a right-invariant metric which
means that
$d_\tau(\sigma_1,\sigma_2)=d_\tau(\sigma_1\sigma,\sigma_2\sigma)$ for
any $\sigma,\sigma_1,\sigma_2\in \FS$ where the operation is the usual
multiplication of permutations. Therefore, we can define the weight of the permutation
$\sigma$ as its distance to the
identity permutation $e=(1,2,\dots,n).$
%
%Also, from the
%definition of the metric $d_\tau$, it follows,
%\begin{equation}
%d_\tau(\sigma_1,\sigma_2) = d_\tau(\sigma_1^{-1},\sigma_2^{-1}).
%\label{inv_metric}
%\end{equation}

Because of the invariance, the graph
whose vertices are indexed by the permutations and edges connect
permutations one Kendall step apart, is regular of degree $n-1.$ At
the same time it is not distance-regular, and so the machinery of
algebraic combinatorics does not apply to the analysis of code
structure. The diameter of the space $X_n$ equals $N\triangleq\binom n2$
and is
realized by pairs of opposite permutations such as $(1,2,3,4)$ and
$(4,3,2,1).$

The main tool to study properties of $d_\tau$ is provided by the
inversion vector of the permutation.  An inversion in a permutation
$\sigma\in \FS$ is a pair $(\sigma(i),\sigma(j))$ such that $i<j$ and
$\sigma(i)>\sigma(j).$
%As a consequence of the invariance property, we
%can define the weight of a permutation $\sigma$ as its distance to the
%identity permutation $e=(1,2,\dots,n).$
It is easy to see that
$d_\tau(\sigma,e)=I(\sigma)$, the total number of inversions in
$\sigma.$ Therefore, for any two permutations $\sigma_1, \sigma_2$ we
have
$d_\tau(\sigma_1,\sigma_2)=I(\sigma_2\sigma_1^{-1})=I(\sigma_1\sigma_2^{-1}).$
In other words,
$$
d_{\tau}(\sigma,\pi) = |\{(i,j)\in [n]^2 : i\ne j, \pi(i) > \pi(j),
\sigma(i) < \sigma(j)\}|.
$$

To a permutation $\sigma\in \FS$ we associate an {\em inversion
  vector} $ \bfx_\sigma\in
G_n\triangleq\integers_2\times\dots\times\integers_n, $ where
$\bfx_\sigma(i)=|\{j: j<i+1, \sigma(j)>\sigma(i+1)\}|, i=1,\dots,n-1$
and $\integers_m$
is the set of integers modulo $m$.  It is well known that the mapping
from permutations to the space of inversion vectors is one-to-one, and
any permutation can be easily reconstructed from its inversion vector.
Moreover,
\begin{equation}
   \label{inversion_1} I(\sigma) =  \sum_{i=1}^{n-1}x_{\sigma}(i).
\end{equation}
For the type of errors that we consider below we introduce the
following $\ell_1$ distance function on $G_n:$
    \begin{equation}\label{eq:d}
    d(\bfx,\bfy)=\sum_{i=1}^{n-1}|x(i)-y(i)|    \qquad(\bfx,\bfy\in G_n)
   \end{equation}
   where the computations are performed over the integers, and write
   $\|\bfx\|$ for the corresponding weight function (this is not a
   properly defined norm because $G_n$ is not a linear
   space)\footnote{This metric is reminiscent of the {\em
       asymmetric distance} function on the set of integer $n$-tuples,
     $\rho(\bfx,\bfy)= \Big|\sum_{i=1}^{n-1} (x(i)-y(i))\Big|$
     \cite{DP1981}.}.  For instance, let $\sigma_1=(2,1,4,3), \sigma_2=(2,3,4,1),$
   then $x_{\sigma_1}=101,x_{\sigma_2}=003.$ To compute the distance
   $d_\tau(\sigma_1,\sigma_2)$ we find
    $$
     I(\sigma_2\sigma_1^{-1})=I((1,4,3,2))=\|(0,1,2)\|=3.
   $$

Observe that the mapping $\sigma\to\bfx_\sigma$
   is a weight-preserving bijection between $X_n$ and the set
   $G_n$.
At the same time, since the groups $\FS$ and $G_n$ are not isomorphic (one
is commutative
while the other is not), this mapping is not distance-preserving.
However, a weaker property is true, namely,
   \begin{equation}\label{eq:dd}
     d_\tau(\sigma_1,\sigma_2)\ge d(\bfx_{\sigma_1},\bfx_{\sigma_2}).
   \end{equation}
%Indeed, transposing two neighboring entries of a permutation $\sigma$
%changes the inversion count $I(\sigma)$ by one, so the mapping $X_n\to
%G_n$ preserves distances to the identity permutation.
Indeed, if the Kendall distance between two permutations
is $1$, then the $\ell_1$ distance between the corresponding two
inversion vectors is $1$ as well. The converse is not necessarily true.

From (\ref{eq:dd}), if there exists a code in $G_n$ with $\ell_1$ distance $d$ then there exists a code
of the same size in $X_n$ with Kendall distance at least $d.$

Another embedding of $X_n$ is given by mapping each permutation
to a binary $ N $-dimensional vector $\bfa$ whose coordinates are indexed
by the pairs $(i,j)\subset[n]^2, i<j,$ and $a_{(i,j)}=1$ if the pair $(i,j)$
is an inversion and $a_{(i,j)}=0$ otherwise. Clearly the Hamming weight
of $\bfa$ equals $I(\sigma),$
and so this mapping is an isometry between $X_n$ and a subset of
the Hamming space $\cH_N$. This mapping was first considered in \cite{CR1970}.

\section{Bounds on the size of rank permutation codes}
An $(n,M,d)$ code $\cC\subset X_n$ is a set of $M$ permutations in which
any two distinct permutations are at least $d$ distance units apart.
Let $A(n,d)$ be the maximum size of the code in $X_n$ with distance $d$.
For the purposes of asymptotic analysis we define the rate of a code $\cC\subset X_n$ of size $M$ as
   $
    R(\cC)=\frac {\ln M}{\ln n!}.
  $
Let
  $$
    \sC(d)=\lim_{n\to\infty}\frac {\ln A(n,d)}{\ln n!}
  $$
be the capacity of rank permutation codes of distance $d$ (our proof of Theorem
\ref{thm:bounds} will imply that the limit exists).
The main result of this section is given in the following theorem
whose proof is given in Sections \ref{sec:spb} and \ref{sec:l1} below.
\begin{theorem} \label{thm:bounds}
   \begin{equation}\label{eq:bounds}
   \sC(d)=\begin{cases} 1 &\text{if } d=O(n)\\
     1-\epsilon &\text{if } d=\Theta(n^{1+\epsilon}), \;0<\epsilon<1\\
     0  &\text{if } d=\Theta(n^2).\end{cases}
  \end{equation}
 \end{theorem}
{\em Remark.} As will be seen from the proof, the equality
$\sC(d)=1-\epsilon$ holds under a slightly weaker condition, namely,
  $
   d={n^{1+\epsilon}}\alpha(n),
  $
where $\alpha(n)$ grows slower than any positive power of $n$.

\subsection{A Singleton bound}
\begin{theorem} Let $d>n-1,$ then
\begin{equation}
\label{singleton} A(n,d) \le \big\lfloor{\nicefrac32+\sqrt{n(n-1)
-2d+\nicefrac14}}\big\rfloor!.
\end{equation}
\end{theorem}
\begin{IEEEproof} Let $\cC$ be an $(n,M,d)$ code. Since the metric $d_{\tau}$ is right
invariant, we can assume that $\cC$ contains the identity permutation $e.$

Let $k \le n$ and let $\cC_k\in {\mathfrak S}_k$ be a code derived from $\cC$ in the
following way. Let $\phi_k: \FS \to {\mathfrak S}_k$ be a mapping that acts on
$\sigma$ by deleting elements $k+1,\dots, n$ from it.
Thus,
%where $\phi_k(\sigma) = \sigma_k$ such that $\forall i,j \in [k]$,
%$\sigma_k(i) < \sigma_k(j)$ if and only if $\sigma(i) \le \sigma(j).$
$\phi_k(\sigma)$ is a permutation on $k$ elements that
maintains the relative positions of the elements of $[k]$ given by
$\sigma$.

Let $k$ be the greatest number such that $\phi_k$ is not injective.
Then $\phi_{k+1}$ is injective, and $M\le (k+1)!$.
Suppose that permutations $\sigma_1, \sigma_2 \in \FS$ are such that $\phi_k(\sigma_1) =
\phi_k(\sigma_2)$. Because of the last equality, none of the
first $k$ entries of the permutation
$\sigma_2\sigma_1^{-1}$ contain pairs that form inversions. Therefore,
    $$
       d\le d_{\tau}(\sigma_1,\sigma_2) \le \binom n2-\binom k2.
   $$
This gives
  $$
  k \le
\frac{1+\sqrt{4n(n-1) -8d+1}}{2},
  $$
which proves inequality (\ref{singleton}). This estimate is nontrivial
 if $\frac32+\sqrt{n(n-1)
-2d+\nicefrac14}<n$ which is equivalent to the condition $d>n-1.$
\end{IEEEproof}

To gain an insight into this bound, let $d=\delta N $.
Using the inequality $m!\le (m/2)^m$ in (\ref{singleton}),
we obtain the asymptotic inequality
 $$
   A(n,d)\le \exp(n(\ln n)\sqrt{1-\delta}(1+c(\ln n)^{-1})),
  $$
%This conforms with the Singleton bound in the Hamming space which gives the
%asymptotic estimate $\exp(n(1-\delta))$ for the size of a code of
%Hamming distance $\delta n.$
%The square root in our estimate is due to the fact that we normalize $d$ by
%$n^2$ rather than by $n$.
where the constant $c$ does not depend on $n$.
As we will show in the next section, the $\sqrt{1-\delta}$
in this bound can in fact be
improved to a quantity that decays as $(\ln n)^{-1}$ as $n$ grows.

\subsection{Sphere packing bounds} \label{sec:spb}
Sphere packing bounds on codes in the Kendall space $X_n$ are related to
the count of inversions in permutations. In this section we discuss several
classic and new results in this area, showing that they imply the asymptotic
scaling order of $\sC(d)$ for very small or very large values of $d$.

Denote by $B_r=B_r(X_n)$ the ball of radius $r$ in $X_n$. Clearly,
   \begin{equation}
      \frac{n!}{|B_{2r}|}\le A(n,2r+1)\le \frac{n!}{|B_r|}.\label{eq:sphere}
   \end{equation}
   The embeddings of $X_n$ into other metric spaces
    can be used to derive estimates of $A(n,d)$ based
   on these inequalities.  In particular, estimating the volume of the
   $\ell_1$-metric ball in $H_n=\{1,\ldots,n\}^n$ and using (\ref{eq:dD}),
    both lower and
   upper bounds will follow from the embedding of $X_n$ in the
   space $H_n$.

Let  $K_n(k)=|\{\sigma\in\FS: I_n(\sigma)=k\}|$ be the number of permutations
with $k$ inversions. By (\ref{inversion_1}), $K_n(k)$ is the number of solutions
of the equation
  $$
    \sum_{i=1}^{n-1}x_i=k, \quad\text{where }x_i\in \integers_{i+1}.
  $$
Then clearly $K_n(k)=0$ for $k> N $ and
  $$
    K_n(k)=K_n\Big( N -k\Big) \quad\text{for }0\le k\le \frac12 N .
  $$
  The number of inversions in a random permutation is asymptotically
  Gaussian with mean $\frac12 N $ and variance
  $\frac{2n^3+3n^2-5n}{72}\approx \frac{n^3}{36}$,
  \cite[p.257]{fel68}. This suggests that codes with distance greater than
$\frac12N$ cannot have large size. We show that this is indeed the case in
Sect.~\ref{sec:H}.

The generating function for the numbers
  $K_n(k)$ has the form
   \begin{equation}
\label{generating} K(z)=\sum_{k=0}^{\infty} K_n(k)z^k
 %= \prod_{i=1}^{n-1}(1+z +z^2 +\ldots+z^{i})
  =  \prod_{i=1}^{n}\frac{1-z^i}{1-z}.
    \end{equation}
For $1\le k\le n$ the number of permutations with $k$ inversions can be
found explicitly \cite{K1973}:
  \begin{multline}
  K_n(k)= {{n+k-2} \choose k} - {{n+k-3} \choose {k-2}} \\+ \sum_{j\ge 2}
(-1)^j \Big[{{n+k-u_j-1} \choose {k-u_j}} +{{n+k-u_j-j-1} \choose
{k-u_j-j}}\Big]\label{eq:vol1},
  \end{multline}
where $u_j=(3j^2-j)/2$ and the summation extends for as long as the binomial coefficients
are positive (it contains about $1.6\sqrt k$ terms).

For $1\le k\le n$ the expression for $K_n(k)$ is given above. In
particular, it implies that $|B_1|=n,$ and $A(n,3)\le (n-1)!.$ As
shown in \cite{M2001}, for $n=k+m, m\to\infty, k\ge 0$
    \begin{equation}\label{eq:vol}
      K_{n}(k)=(0.289\ldots) \frac{2^{m+n-1}}{\sqrt{\pi m}}(1+O(m^{-1})).
    \end{equation}
The case of $k>n$ is much more difficult to analyze. An obvious
route for finding asymptotic approximation of $K_n(k)$ is to start
with the integral representation of the coefficients of $K(z)$
(\ref{generating}). Namely, since $K(z)$ converges for every $z$
in the finite plane, we can write
   $$
    K_n(k)=\frac 1{2\pi i}\oint_{C} \prod_{\ell=1}^n
            \Big(\frac{1-z^\ell}{1-z}\Big)z^{-k-1}dz.
  $$
where $C$ is a circle around the origin.
\remove{In particular, for $|z|=1$ we obtain
   \begin{eqnarray}\label{eq:Knk}
K_n(k) &=& \frac{1}{2\pi}\int_{-\pi}^{\pi} \prod_{\ell=1}^{n}
\frac{1-e^{i \ell \omega }}{1- e^{i \omega }}e^{-ik\omega}d\omega\\
&=& \frac{2}{\pi} \int_{0}^{\pi/2} \cos\left(x\left({n \choose 2}-
2k\right)\right)\prod_{\ell=1}^{n}  \frac{\sin(x \ell)}{\ell\sin{x}} dx.\nonumber
   \end{eqnarray}
}
\remove{A recent paper \cite{C2000} analyzes this integral;
 however, since it pursues a detailed asymptotic expansion for
 $K_n(k)$, it ends up with a narrow interval in which this
     expansion applies. To be more specific this paper provides the
     asymptotic expansion of the number $K_n(k)$ for $|k-\frac12 N
     |\le \sqrt{n^3\ln n}/6.$}
Asymptotic analysis of this expression involves saddle point calculations
and is rather involved \cite{lou03}.
The next theorem is a combination of results of Margolius \cite{M2001} and
Louchard and Prodinger \cite{lou03}, stated here
in the form suitable for our context.
\begin{theorem}\label{thm:sphere}
There exist constants $c_1$ and $c_2$ such that
     \begin{align*}
     K_n(k)  &\le \exp(c_1n) & &\text{if } k = O(n), \\
     K_n(k)&= n!/\exp(c_2n) & &\text{if } k = \Theta(n^2).
     \end{align*}
\end{theorem}
The implicit constants in this theorem can be found in cited references.

 From this theorem and inequalities (\ref{eq:sphere}), we obtain the two
boundary cases of the expression for $\sC(d)$ in (\ref{eq:bounds}).

\subsection{Bounds from embedding in the $\ell_1$ space}\label{sec:l1}

In this section we prove the remaining case of Theorem \ref{thm:bounds}.
Our idea is to derive bounds on $\sC(d)$
by relating the Kendall metric to the $\ell_1$ metric on $\FS.$
 From the results of Diaconis and Graham \cite{DG1977},
\begin{equation}\label{eq:dD}
\half D(\sigma_1,\sigma_2)\le d_{\tau}(\sigma_1^{-1},\sigma_2^{-1}) \le D(\sigma_1,\sigma_2).
\end{equation}
where $D(\sigma_1,\sigma_2)=\sum_{i=1}^n |\sigma_1(i)-\sigma_2(i)|.$
Therefore, existence of any code $\cC\subset \FS$ with Kendall distance $d$ must
imply existence of a code $\cC' = \{\sigma^{-1}: \sigma \in \cC\}$
of same size that have $\ell_1$ distance
 at least $d$. On the other hand existence of any code $\cC\subset \FS$ with $\ell_1$ distance $d$
implies the code $\cC' = \{\sigma^{-1}: \sigma \in \cC\}$
 will have Kendall distance at least $d/2.$

{\em Remark.} Define $T(\sigma_1,\sigma_2)$ to be the number of inversions
of (not necessarily adjacent) symbols needed to change $\sigma_1$ into
$\sigma_2$.  Paper \cite{DG1977} in fact shows that
  $$
    d_\tau(\sigma_1^{-1},\sigma_2^{-1})\le D(\sigma_1,\sigma_2) -T(\sigma_1,\sigma_2)
  $$
which is a stronger inequality than the one given above.
We however will not use it in the derivations below.

\begin{proposition} Let $B_r(H_n,\bfx)$ be the  metric
ball of radius $r$ with center at $\bfx$ in the space
$H_n=\{1,2,\dots,n\}^n$ with the $\ell_1$ metric. Then the maximum size
of a code in $X_n$ with distance $d$ satisfies
   $$
   \frac{n!}{\max_{\bsx\in H_n}|B_{2d-1}(H_n,\bfx)|}
 \le A(n,d) \le \frac{n^n}{\min_{\bsx \in H_n}|B_t(H_n,\bsx)|},
   $$
where  $t=\lfloor(d-1)/2\rfloor.$
\end{proposition}
\begin{IEEEproof}
 Under the trivial embedding
$\FS\to H_n$ the $\ell_1$ distance does not change, so any code
$\cC$ in $\FS$ with $\ell_1$ distance $d$ is also a code in $H_n$
with the same distance and as such, must satisfy the Hamming bound.
Together with (\ref{eq:dD}) this gives the upper bound of our statement.

Turning to the lower bound, let us perform the standard
``Gilbert procedure'' in the space of permutations with
respect to the $\ell_1$ distance, aiming for a code $\cD$ with $\ell_1$
distance $m$.
The resulting code satisfies
  $$
   |\cD| \max_{\sigma\in \FS} |B_{m-1}(\FS,\sigma)|\ge n!.
  $$
Since $|B_r(H_n,\sigma)|\ge |B_r(\FS,\sigma)|,$ we can replace the
volume in $\FS$ with the volume in $H_n$ in the last inequality.
In the space $X_n$, the code $\cD' = \{\sigma^{-1}: \sigma \in \cD\}$ will then have
Kendall distance at least $m/2.$
\end{IEEEproof}

Below we consider only spheres in the space $H_n$ and omit the
reference to it from the notation $B_r(H_n,\cdot)$.
\begin{lemma}
Let $\textbf{1} = (1,1,\ldots,1) \in H_n.$ Then for any $\bfz,\bsy \in H_n$,
$$
2^{-n}|B_r(\bfz)|\le |B_{r}(\textbf{1})| \le |B_{r}(\bsy)|.
$$
\end{lemma}
\begin{IEEEproof}
%To prove this lemma we will construct an injective mapping from
%$B_{r}(\textbf{1})$ to $B_{r}(\bsy)$ for any $\bsy \in H_n$.
  Suppose that $\bsx = (x_1,x_2,\ldots,x_n) \in B_{r}(\textbf{1})$ and
  $\textbf{1} \ne \bsy = (y_1,y_2,\ldots,y_n) \in H_n$. Consider the
  mapping $\zeta: B_{r}(\textbf{1}) \to B_{r}(\bsy)$ where $\bsx
  \mapsto \bsu,$ where $\bsu= (u_1,u_2,\ldots,u_n)$ is given by
\begin{equation*}
    u_i = \begin{cases} y_i + (x_i -1) & \text{if } y_i + (x_i -1) \le n \\
     n-(x_i-1) & \text{if } y_i + (x_i -1) > n.
     \end{cases}
  \end{equation*}
Clearly $\bsu \in H_n$ and $x_1 -1 \ge |u_i -y_i|$ for $i=1,\ldots,n,$
so every point within distance $r$ of $\textbf{1}$ is sent to a point
within distance $r$ of $\bsy.$ Furthermore, this mapping is injective
because if $\bfx_1, \bfx_2$ are two distinct points in $B_{r}(\textbf{1})$
then their images can coincide only if in some coordinates
  $$
   y_i+(x_{1,i}-1)=n-(x_{2,i}-1).
  $$
However, the left-hand side of this equality is $\ge y_i$ while the
right-hand side is $<y_i$ by definition of $u_i.$
This proves the right inequality.

To prove the lower bound, write $B_r(\bsz)$ as $\bfz+D_r(\bsz)$,
where $D_r(\bsz)$ is the set of differences:
\begin{multline*}
D_r(\bsz)=\{\bsu\in \integers^n: |u_i|\le n-1, 1\le i\le n;\;\sum_{i=1}^n |u_i| \le r
\\\text{ and } \bsz+\bsu \in H_n \}.
\end{multline*}
Writing $B_r(\textbf{1})$ in the same way as $\textbf{1}+D_r^{+},$ we have
$$
D_r^+ = \{\bsu \in \integers^n: 0\le u_i\le n- 1;\;\sum_{i=1}^n |u_i| \le r\}.
$$
By taking the absolute values of the coordinates, any point in
$D_r(\bsz)$ is sent to a point in $D_r^+$, and no more than $2^n$
points have the same image under this mapping.
This proves our claim.
\end{IEEEproof}

These arguments give rise to the next proposition.
\begin{proposition}
\begin{align}
  \frac {n!}{2^n \sum_{r=0}^{2d-1} Q(n,r)} \le A(n,d)
   \le \frac{n^n}{\sum_{r=0}^t
Q(n,r)}, \label{eq:spb}
%\le \frac{n^n}{Q(n,t)}
\end{align}
where
  $$Q(n,r) = \sum_{i\ge 0} (-1)^i K_{n,r}(i)$$
and
$ K_{n,r}(i) = {n \choose i}{n+r -ni -1 \choose r-ni}$.
\end{proposition}
This claim is almost obvious because,
by the previous lemma,
  $$\frac{n!}{2^n|B_{2d-1}(\textbf{1})|}\le A(n,d) \le \frac{n^n}{|B_t(\textbf{1})|}
   $$
Next,
  $$
  |B_s (\textbf{1})| =\sum_{r=0}^s
Q(n,r),
  $$
where $Q(n,r)$
is the number of integer solutions of the equation
$$
x_1 +x_2 + \ldots + x_n =r ,
$$
where $0 \le x_i \le n-1,$ $1\le i\le n.$
The expression for $Q(n,r)$ given in the statement is well known
(e.g.,\cite[p.1037]{GS1995}).

Expression (\ref{eq:spb}) gives little insight into the behavior
of the bound. In the remainder of this section we estimate the asymptotic
behavior of this bound and derive an estimate of the code capacity.

\begin{lemma}\label{lemma:Qnr}
Suppose that $r < n^2/\ln n.$ Then
$$
{n+r-1 \choose r}-n\binom{r-1}{r-n}\le Q(n,r) \le {n+r-1 \choose r}.
$$
\end{lemma}
\begin{IEEEproof}
Let $S(n,j)=\sum_{i\ge j}(-1)^i K_{n,r}(i).$
The lemma will follow if we prove that
  \begin{equation}\label{eq:S}
    S(n,1)< 0 \text{ and } S(n,2)>0.
  \end{equation}
\remove{Let us write the definition of $Q(n,r)$ as follows:
  $$
  Q(n,r)={n+r-1 \choose r} +\sum_{i\ge 1}(-1)^i K_{n,r}(i).
  $$
We need to prove that the last sum is negative for large $n$.}
Under the assumption on $r$ we have
  \begin{align*}
  \frac{\binom {r+n-n(i+1)-1}{r-n(i+1)}}{\binom{r+n-ni-1}{r-ni}}
  &=\prod_{j=1}^{n-1}\frac{r-ni-n+j}{r-ni+j}\\
  &=\prod_{j=1}^{n-1}\Big(1-\frac n{r-ni+j}\Big)\\
&\le \Big(1-\frac n{r-n(i-1)-1}\Big)^{n-1}
\end{align*}\begin{align*}
  &\le e^{-\frac{n(n-1)}{r-n(i-1)-1}}\\
  &\le n^{-\frac{n-1}n}\\
  &\le\frac{\sqrt 2}n.
  \end{align*}
Thus for $i\ge 1$
  $$
   \frac{K_{n,r}(i+1)}{K_{n,r}(i)}\le \frac{n-i}{i+1}\frac{\sqrt 2}n<1.
  $$
Therefore $-K_{n,r}(2m-1)+K_{n,r}(2m)<0$ for all $m.$
Since the sum $S(n,1)$ starts with a negative term and the sum $S(n,2)$
with a positive one, the required inequalities in (\ref{eq:S}) follow.
\end{IEEEproof}

 From the foregoing arguments we now have the following explicit
bounds on $A(n,d):$
    \begin{equation}\label{eq:upper}
\frac{n!}{2^n\binom {n+2d-1}{2d-1}}\le A(n,d) \le
\frac{n^n}{\sum_{r=0}^t \big(\binom{n+r-1}{r}-n\binom{r-1}{r-n}\big)}.
\end{equation}
Here the right part is obvious and for the left inequality we used
(\ref{eq:spb}), Lemma \ref{lemma:Qnr},
and the identity $\sum_{i\le n}\binom{s+i}i=\binom{s+n+1}n.$

Now we are ready to complete the proof of Theorem \ref{thm:bounds}.
Assume that $d=\Theta(n^{1+\epsilon})$ for some $0\le \epsilon < 1.$
The two boundary cases of (\ref{eq:bounds}) were established in the
previous section. Let us prove the middle equality.
 From (\ref{eq:upper}),
  $$
A(n,d)
       \le \frac{n^n}{\binom{n+t-1}{n-1}-n\binom{t-1}{t-n}}
  $$
To estimate the denominator, write
  \begin{align*}
   \binom{n+t-1}{n-1}&=\binom{t-1}{t-n}\prod_{j=1}^{n-1}\Big(1+\frac n{t-j}\Big)
   \\&>\binom{t-1}{t-n}\Big(1+\frac nt\Big)^{n-1}\\
   &>n \binom{t-1}{t-n}e^{(n-1)(\frac{n}{t}-\frac12(\frac{n}{t})^2)-\ln{n}}\\
   &= n \binom{t-1}{t-n}e^{\Theta(n^{1-\epsilon})}
  \end{align*}
(because of $   \ln(1+n/t)>(n/t)-\frac12(n/t)^2,    $ for $n/t <1.$)
So starting with some $n$ we can estimate the denominator below by
   $\half \binom{n+t-1}{n-1}.$
 Therefore,
  $$
    A(n,d)\le \frac{2n^n}{\binom{n+t-1}t}\le
\frac{2n^n(n-1)^{n-1}}{(n+t-1)^{n-1}}.$$
%       &\le \frac{n^n(n-1)^{n-1}}{(n+t-1)^{n-1}},
%where we used the fact that ${ n\choose k} \ge (n/k)^k.$
Next
   $$
    \frac{\ln{A(n,\Theta(n^{1+\epsilon}))} }{n\ln n}
  \le 2-(1+\epsilon) + o(1) = 1-\epsilon +o(1).
   $$
On the other hand, using
  $$
  \binom{n+2d-1}{2d-1}\le\Big(\frac{(n+2d)e}n\Big)^n<(2e)^n\Theta(
   n^{n\epsilon })
  $$
 and $n!> (n/3)^n,$
we obtain from (\ref{eq:upper})
  $$
   A(n,d)\ge \frac{n^{n}}{(12 e)^n \Theta(n^{n\epsilon})}.
  $$
Taking the logarithms and the limit, we find that $\sC(d)\ge 1-\epsilon.$
  This completes the proof of Theorem \ref{thm:bounds}.

\subsection{Bounds from embedding in Hamming space}\label{sec:H}
Since the embedding of $X_n$ into the Hamming space $\cH_N$ of dimension
$N=\binom n2$ is isometric, the known results for codes correcting
Hamming errors can be used to derive estimates and constructions for
codes in the Kendall space.  In particular, the known bounds on codes
in the Hamming space can be rewritten with respect to the space $X_n.$
For instance, the Plotkin bound implies that
   $$
      A(n,d)\le 2d/(2d-N)
   $$
   and thus any code $\cC\subset X_n$ with distance greater than the
   average (i.e., $\frac12N$) satisfies $|\cC|=O(N).$

%Results of this kind are unlikely to be very good in general because the image of
%$X_n$ in $\cH_N$ forms a rather small subset of it.
Given the image of a code $\cC\subset X_n$ in $\cH_N$ it is easy to
reconstruct the code $\cC$ itself.
Indeed, it is immediate to find the inversion vector of a permutation $\sigma$
given the image of $\sigma$ in $\cH_N,$ and then to recover
$\sigma$ from its inversion vector.

Of course, not every code in $\cH_N$ will have a code in $X_n$
corresponding to it.
The next simple proposition shows that nevertheless, binary codes in $\cH_N$
can be used to claim existence of good rank permutation codes.
\begin{proposition} Suppose that there exists a binary linear
$[{n \choose 2},k,d]$ code $\cA.$
Then there exists an $(n,\ge\!\!\frac{n!}{2^{ N -k}},d)$
rank permutation code.
\end{proposition}
\begin{IEEEproof}
One of the $2^{N-k}$ cosets of $\cA$ in $\cH_N$ must contain at least
${n!}/{2^{ N -k}}$ vectors that map back to valid permutations.
\end{IEEEproof}

For example, let us assume that the value $N$ is such that there
exists a $t$-error-correcting binary BCH code of length $N$ (if
not, we can add zeros to a shorter BCH code). Its dimension is at
least $N - t\log_2( N +1).$ This shows the existence of a
$t$-error-correcting rank permutation code of size $\frac{n!}{( N
  +1)^t}= \frac{n!}{O(n^{2t})}.$

On the other hand, by the sphere packing bound the size of a
$t$-error-correcting code in $X_n$ is at most $M\le
O(\frac{n!}{n^{t}}).$ Thus, using the embedding $X_n\to \cH_N$ we are
not able to close a gap between the existence results and the upper
bounds.  In the next section we use a different method to construct
codes that achieve the sphere packing bound to within a constant
factor for any given $t$.

\section{Towards optimal $t$-error-correcting codes}
The representation of permutations by inversion vectors provides a way
to construct error-correcting rank permutation codes. In this section
we construct codes in the $\ell_1$ space of inversion vectors $G_n$
and claim the existence of rank permutation codes by the inequality on
the code distances (\ref{eq:dd}).

We begin with constructing codes over the integers that correct
additive errors. Once this is accomplished, we will be able to claim
existence of good rank permutation codes.  Let $A$ be some subset of
$\integers$ and let $A^L$ be the space of $L$-tuples of integers from $A$
equipped with the $\ell_1$ distance (\ref{eq:d}). A code $\cD\subset
A^L$ is said to correct $t$ additive errors if for any two distinct
code vectors $\bfx,\bfy$ and any $\bfe_1,\bfe_2\in \integers^L,$ both
of weight at most $t$,
  $$
  \bfx+\bfe_1\ne\bfy+\bfe_2.
  $$
  We assume that $A$ and $t$ are such that $\cD$ is well defined: for
  instance, below we will take $A=\integers_s$ where $s$ is some
  integer sufficiently large compared to $t$.

  If in the above definition $\bfe_i\ge 0$ for all $i$, the code is
  said to correct $t$ {\em asymmetric} errors. However below we need
  to consider the general case, focusing on a particular way of
  constructing codes which we proceed to describe.
\begin{definition}
Let $m\ge L$ and let $h_1,\dots,h_L, 0<h_i<m, i=1,\dots,L$ be a set of integers.
Define the code as follows:
   \begin{equation}\label{eq:code}
     \cC=\Big\{\bfx\in A^L\Big| \sum_{i=1}^L h_ix_i\equiv 0 \mod{m}\Big\}.
   \end{equation}
\end{definition}
This code construction was first proposed by Varshamov and Tenenholtz
\cite{var65} for correction of one asymmetric error (it was
rediscovered later by Constantin and Rao \cite{con79} and, in a
slightly different context, by Golomb and Welch
\cite{gol70}). Generalizations to more that one error as well as to
arbitrary finite groups were studied by Varshamov \cite{V1973},
Delsarte and Piret \cite{DP1981}, and others; however, these works
dealt with asymmetric errors. Below we extend this construction to the
symmetric case.
 \begin{proposition}\label{prop:syndrome}
   The code $\cC$ defined in (\ref{eq:code}) corrects $t$ additive
   errors if and only if for all $\bfe\in \integers^L, \|\bfe\|\le t$
   the sums $\sum_{i=1}^L e_i h_i$ are all distinct and nonzero modulo
   $m$.
    \end{proposition}
    This proposition is obvious as it amounts to saying that all the
    syndromes of error vectors of weight up to $t$ are different and
    nonzero.

We will need the following theorem of Bose and Chowla \cite{BC1962}.
\begin{theorem}
\label{bosechowla} (Bose and Chowla) Let $q$ be a power of a prime and $m =
(q^{t+1} -1)/(q-1).$ There exist $q+1$ integers
$j_0 =0, j_1,\ldots, j_q$ in $\integers_m$ such  that the sums
   $$
   j_{i_1}+j_{i_2}+\ldots+ j_{i_t}  \quad(0\le i_1\le i_2 \le \ldots \le i_t\le q)
   $$
are all different modulo $m.$
\end{theorem}

This theorem provides a way of constructing an asymmetric
$t$ additive error-correcting code of length $q$.  This is because for any
error vector $\bse$ with $||\bse|| \le t<m$ such that $e_i \ge 0$, the
sums $\sum_{i=1}^{q} e_i j_i$ involve at most $t$ of the numbers $j_i$
and thus are all different. This theorem was previously used to
construct codes in a different context in \cite{gra80} as well as in
some later works.

\begin{theorem} For $1\le i\le q+1$ let
\begin{equation*}
   h_{i} =\begin{cases}  j_{i-1} +\frac{t-1}{2}m  &\text{for $t$ odd}\\
         j_{i-1} +\frac{t}{2}m &\text{for $t$ even}
         \end{cases}
\end{equation*}
where the numbers $j_i$ are given by the Bose-Chowla theorem.
Let $m_t=t(t+1)m$ if $t$ is odd and $m_t=t(t+2)m$ if $t$ is even.
For all $\bse \in \integers^{q+1}$ such that $||\bse|| \le t$ the sums
$\sum_{i=1}^{q+1} e_i h_i$ are all distinct and
nonzero modulo $m_t.$
\end{theorem}
\begin{IEEEproof} Let $t$ be odd and let
$H=\{0,h_1,\ldots,h_{q+1}\}.$ Observe that
  \begin{equation}\label{eq:h}
  (t-1)m/2 \le h_i < (t+1)m/2.
  \end{equation}

(i) For any $k_1\le k_2\le \ldots\le k_t \in H$, the sums $\sum_{i=1}^t
k_i$ are all distinct modulo $m$ and therefore also modulo $m_t.$ These sums are also
nonzero modulo $m$ except for the case when all the $k_i$'s are $0$.

(ii) Moreover, for any $k_1\le k_2\ldots\le k_{2t} \in H$, the sum
  $$
  \sum_{i=1}^{2t} k_i <  m_t,
  $$
and is therefore nonzero modulo $m_t.$

(iii) Finally, for any $0<k_1, k_2, \ldots, k_{2t} \in H$ and any $r<t,$
  \begin{align}
  \sum_{i=2t-r+1}^{2t} k_i&< r\frac{t+1}{2}m\le (2t-r)\frac{t-1}{2}m\nonumber\\
    & \le \sum_{i=1}^{2t-r} k_i. \label{eq:k}
  \end{align}

Let us suppose now that there exist nonzero vectors $\bse_1, \bse_2 \in
\mathbb{Z}^{q+1}$ both of weight at most $t$ such that
  \begin{eqnarray*}
  \text{either  (a)}&\sum_{i=1}^{q+1} e_{1i} h_i =0 \mod{m_t}\\ \text{ or (b)}&
  \sum_{i=1}^{q+1} e_{1i} h_i
=\sum_{i=1}^{q+1} e_{2i} h_i \mod{m_t}.
   \end{eqnarray*}

However assuming (a) contradicts property (i). On the other hand if (b)
is true then one of the following two scenarios can happen. In the
first case, $e_{1i}\ge 0$ and $e_{2i}\le 0$ for all $i$ or
$e_{1i}\le 0$ and $e_{2i}\ge 0$ for all $i$. It is easy to see that
this assumption contradicts
property (ii). In all other situations, (b)
contradicts either property (i) or property (iii) above.

The claim for $t$ even is proved in an analogous way. Namely, we will
have
  $$
   tm/2\le h_i\le (t+2)m/2
  $$
and
  $$
  \sum_{i=2t-r+1}^{2t}k_i< r\Big(\frac{t+2}2\Big)m\le (2t-r)\frac {tm}2\le
  \sum_{i=1}^{2t-r} k_i
  $$
instead of (\ref{eq:h}) and (\ref{eq:k}), respectively.
The rest of the proof remains the same.
\end{IEEEproof}

Together with Proposition \ref{prop:syndrome} this theorem implies the
existence of a $t$-error-correcting code $\cC$ of length $q+1$ over %$n-1$ over
the alphabet $\integers_{m_t}$ that corrects $t$ additive errors.
Recall that our goal is to construct a code over
the set of inversion vectors $G_n$ that corrects $t$ additive errors.
At this point
let us set $q+1 = n-1.$
 Note that, $G_n$ is a subset of $\integers_n^{n-1}$
which is a subset of $\integers_{m_t}^{n-1}.$
Since $\cC$ is a group code with respect to addition modulo $m_t$, its
cosets in $\integers_{m_t}^{n-1}$ partition this space into disjoint equal
parts. At least one such coset contains $M\ge n!/m_t$ vectors from
$G_n.$ Invoking (\ref{eq:dd}) we now establish the main result of this
section.
\begin{theorem} Let $m=((n-2)^{t+1}-1)/(n-3),$ where $n-2$ is a power of a
prime. There exists a
  $t$-error-correcting rank permutation code in $\FS$ whose size
  satisfies
  \begin{equation*}
    M\geq\begin{cases} n!/(t(t+1)m) & (t\text{ odd}) \\
     n!/(t(t+2)m) &(t\text{ even}).
     \end{cases}
  \end{equation*}
\end{theorem}
This theorem establishes the existence of codes whose size is of the
same order $O(n!/n^t)$ as given by the sphere packing bound of the
previous section. The loss of a constant multiple of the optimal code
size is due to the fact that we construct codes over the integer
alphabet instead of a more restricted alphabet
$\integers_2\times\ldots \times\integers_n$.

As a final remark, note that the construction is explicit except for
the last step where we claim existence of a large-size code in some
coset of the code $\cC.$

\vspace*{2mm}
{\em Acknowledgment.} The authors are grateful to Gregory Kabatiansky for a
useful discussion of this work.

\balance
\end{document}